\documentclass[sigconf]{acmart}
\usepackage{enumitem}
\usepackage{amsfonts}
\usepackage{multirow}
\usepackage{threeparttable}
\usepackage{arydshln}
\usepackage{adjustbox}
\usepackage{graphicx}
\usepackage{algorithm}
\usepackage{algpseudocode}
\usepackage{listings}
\usepackage{xcolor}
\AtBeginDocument{%
  }

\copyrightyear{2025}
\acmYear{2025}
\setcopyright{acmlicensed}
\acmDOI{10.1145/3701716.3715208}
\acmConference[WWW Companion '25] {Companion of the 16th ACM/SPEC International Conference on Performance Engineering}{April 28-May 2, 2025}{Sydney, NSW, Australia.}
\acmBooktitle{Companion of the 16th ACM/SPEC International Conference on Performance Engineering (WWW Companion '25), April 28-May 2, 2025, Sydney, NSW, Australia}
\acmISBN{979-8-4007-1331-6/25/04}

\begin{document}

\title{A Hybrid Cross-Stage Coordination Pre-ranking Model for Online Recommendation Systems}


\author{Binglei Zhao}
\email{zhaobinglei1@jd.com}
\orcid{0000-0002-7334-9452}
\affiliation{%
    \institution{JD.com, Inc}
    \city{Beijing}
    \country{China}
}

\author{Houying Qi}
\email{qihouying@jd.com}
\orcid{0009-0008-7890-5324}
\affiliation{%
    \institution{JD.com, Inc}
    \city{Beijing}
    \country{China}
}

\author{Guang Xu}
\email{xuguang66@jd.com}
\orcid{0009-0008-9969-4810}
\affiliation{%
    \institution{JD.com, Inc}
    \city{Beijing}
    \country{China}
}

\author{Mian Ma}
\email{mamian@jd.com}
\orcid{0000-0001-9969-6623}
\affiliation{%
    \institution{JD.com, Inc}
    \city{Beijing}
    \country{China}
}

\author{Xiwei Zhao}
\email{zhaoxiwei@jd.com}
\orcid{0000-0002-9382-6041}
\affiliation{%
    \institution{JD.com, Inc}
    \city{Beijing}
    \country{China}
}

\author{Feng Mei}
\email{meifeng6@jd.com}
\orcid{0009-0002-9442-2548}
\affiliation{%
    \institution{JD.com, Inc}
    \city{Beijing}
    \country{China}
}

\author{Sulong Xu}
\email{xusulong@jd.com}
\orcid{0000-0003-0345-334X}
\affiliation{%
    \institution{JD.com, Inc}
    \city{Beijing}
    \country{China}
}

\author{Jinghe Hu}
\email{hujinghe@jd.com}
\orcid{0009-0002-1546-5807}
\affiliation{%
    \institution{JD.com, Inc}
    \city{Beijing}
    \country{China}
}

\renewcommand{\shortauthors}{Binglei Zhao, et al.}

\begin{abstract}
Large-scale recommendation systems often adopt cascading architecture consisting of retrieval, pre-ranking, ranking, and re-ranking stages.
With strict latency requirements, pre-ranking utilizes lightweight models to perform a preliminary selection from massive retrieved candidates.
However, recent works focus solely on improving consistency with ranking, relying exclusively on downstream stages.
Since downstream input is derived from the pre-ranking output, they will exacerbate the sample selection bias (SSB) issue and Matthew effect, leading to sub-optimal results.
To address the limitation, we propose a novel \textbf{H}ybrid \textbf{C}ross-Stage \textbf{C}oordination \textbf{P}re-ranking model (\textbf{HCCP}) to integrate information from upstream (retrieval) and downstream (ranking, re-ranking) stages.
Specifically, cross-stage coordination refers to the pre-ranking's adaptability to the entire stream and the role of serving as a more effective bridge between upstream and downstream.
HCCP consists of \textbf{Hybrid Sample Construction} and \textbf{Hybrid Objective Optimization}.
Hybrid sample construction captures multi-level unexposed data from the entire stream and rearranges them to become the optimal guiding "ground truth" for pre-ranking learning.
Hybrid objective optimization contains the joint optimization of consistency and long-tail precision through our proposed Margin InfoNCE loss. 
It is specifically designed to learn from such hybrid unexposed samples, improving the overall performance and mitigating the SSB issue.
The appendix describes a proof of the efficacy of the proposed loss in selecting potential positives.
Extensive offline and online experiments indicate that HCCP outperforms SOTA methods by improving cross-stage coordination.
It contributes up to \textbf{14.9\%} UCVR and \textbf{1.3\%} UCTR in the JD E-commerce recommendation system.
Concerning code privacy, we provide a pseudocode for reference.
\end{abstract}

\vspace{-10mm}

\vspace{-10mm}

\vspace{-10mm}


\maketitle

\vspace{-2mm}
\begin{figure}[h]
  \centering
  \vspace{-6mm}
\includegraphics[width=0.42\textwidth, height=2.3cm]{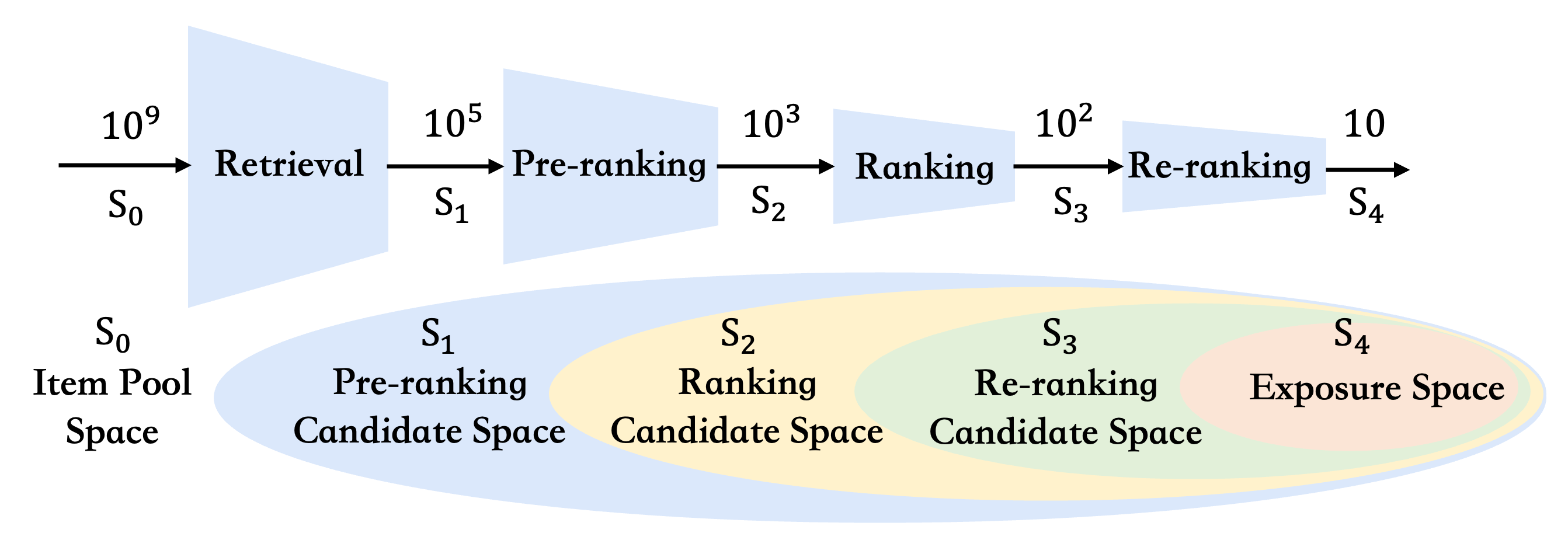}
  \vspace{-3mm}
   \caption{Cascade architecture in recommendation system.}
  \vspace{-3mm}
\label{fig:cascade}
\end{figure}

\vspace{-2mm}
\section{Introduction}
Large-scale industrial recommendation systems typically employ cascade structures, consisting mainly of retrieval, pre-ranking, ranking, and re-ranking stages.
The retrieval stage samples massive items ($S_1$ \textasciitilde{} $10^5$) and then pre-ranking selects a reduced set ($S_2$ \textasciitilde{} $10^3$).
Afterward, they go through ranking and re-ranking stages, and ultimately, about ten items ($S_4$ \textasciitilde{} $10$) are chosen to be exposed, as shown in Figure\ref{fig:cascade}.
Pre-ranking plays a pivotal role in recommendation systems, performing a preliminary selection from large-scale retrieval candidates and distilling an optimal subset for ranking.

To strike an optimal balance between effectiveness and efficiency, pre-ranking typically adopts vector-product-based models\cite{2013dssm,2015lstmdssm}.
Some works design new models\cite{2020cold, 2021fscd,2022inttower} to achieve improvements, which optimize pre-ranking as an independent entity.
Despite advancements, lightweight models lead to an inevitable gap and inconsistency with ranking, which lowers system performance particularly when ranking favors items overlooked by pre-ranking.
Recent studies\cite{2018_rank_distillation, 2022rankflow, 2023_list_distillation,2023rethinking} focus on strict score alignment to approximate ranking ability.
Considering alignment errors, a $\Delta$NDCG-based method\cite{2023_copr_ndcg} is proposed to distinguish relative orders of chunks.
They overlook the accuracy distinction of ranking models for top and tail items when imitating ranking orders, either treating them indistinguishably\cite{2023_copr_ndcg} or outright discarding tail items\cite{2023rethinking}, potentially affecting pre-ranking's accuracy on tail items.
Furthermore, given that downstream input is derived from the pre-ranking output, blindly imitating ranking orders and relying exclusively on downstream will limit the ability of pre-ranking to serve as a bridge connecting retrieval and ranking, which may exacerbate sample selection bias (SSB)\cite{2004ssb} and Matthew effect.
For pre-ranking, selecting high-quality mid-tail items and enhancing long-tail precision is essential.
A fundamental idea is increasing massive unexposed data during training. 
Nevertheless, since user feedback on these data is inherently unknown, directly considering them as negatives will introduce a significant bias. 
The pivotal question addressed in this paper is: How can unexposed data with entire stream information be leveraged to make pre-ranking a more suitable bridge?

We analyze the information gain of unexposed samples in \textbf{\textit{entire stream}}: 1. downstream (ranking, re-ranking) gives an accurate prediction on click-through rate (CTR) and conversion rate (CVR) of top items. 2. upstream (retrieval) offers relevant items with greater efficiency, albeit they may not be exposed.
To this end, we propose an innovative concept of \textbf{\textit{cross-stage coordination}} of pre-ranking, which refers to integrating information from upstream and downstream stages.
We also propose a \textbf{H}ybrid \textbf{C}ross-Stage \textbf{C}oordination \textbf{P}re-ranking method (\textbf{HCCP}) for online recommendation systems.
The crux lies in hybrid sample construction and hybrid objective joint optimization on sampled multi-level data.
\textbf{\textit{Hybrid sample construction}} aims at capturing multi-level unexposed data from the entire stream, such as ranking sequence, pre-ranking sequence, in-batch negatives, and pool sampling negatives. 
The ranking sequence is non-uniformly sampled from downstream and rearranged to integrate with the diversity information of re-ranking and user preferences, which guarantees the "ground truth" is more suitable for pre-ranking learning.
Compared to in-batch and pool sampling negatives, unexposed ranking and pre-ranking sequences are usually regarded as hard negatives.
Various samples serve different learning objectives.
\textbf{\textit{Hybrid objective optimization}} is specifically designed to learn such instances that do not consistently correlate with user feedback.
Hybrid objectives encompass consistency with downstream and long-tail item precision.
For consistency learning, we propose list-wise global and local consistency losses to achieve soft order alignment on ranking sequence, which enables pre-ranking to approximate ranking ability of complex ranking models.
For long-tail precision optimizing, we propose a novel Margin InfoNCE loss and incorporate priors of negative difficulties into contrastive learning, further enhancing discriminative capacity and mitigating SSB issues.
In the appendix, we provide theoretical proof for the Margin InfoNCE loss in distinguishing potential positives from hard negatives. 
HCCP explicitly integrates entire stream data, rather than treating pre-ranking as an isolated stage or focusing solely on consistency.
The contributions are summarized as follows:
\vspace{-5mm}
\begin{itemize} [leftmargin=*]
\item We propose the \textbf{H}ybrid \textbf{C}ross-\textbf{S}tage \textbf{C}oordination \textbf{P}re-ranking model, which explicitly integrates entire stream information and makes pre-ranking a more suitable bridge.
It can migrate to any model with almost no latency increase.


\item We propose a hybrid sample construction module to capture multi-level unexposed data from upstream and downstream and a hybrid objective optimization module to enhance consistency and long-tail precision through a novel Margin InfoNCE loss.


\item Extensive experiments verify HCCP's effectiveness in offline dataset and online system.
It outperforms state-of-the-art baselines, achieving a 14.9\% increase in UCVR and a 1.3\% increase in UCTR in JD E-commerce recommendation system.
\end{itemize}

\begin{figure*}
  \centering
\vspace{-4mm}
  \includegraphics[width=1\textwidth, height=4.5cm]{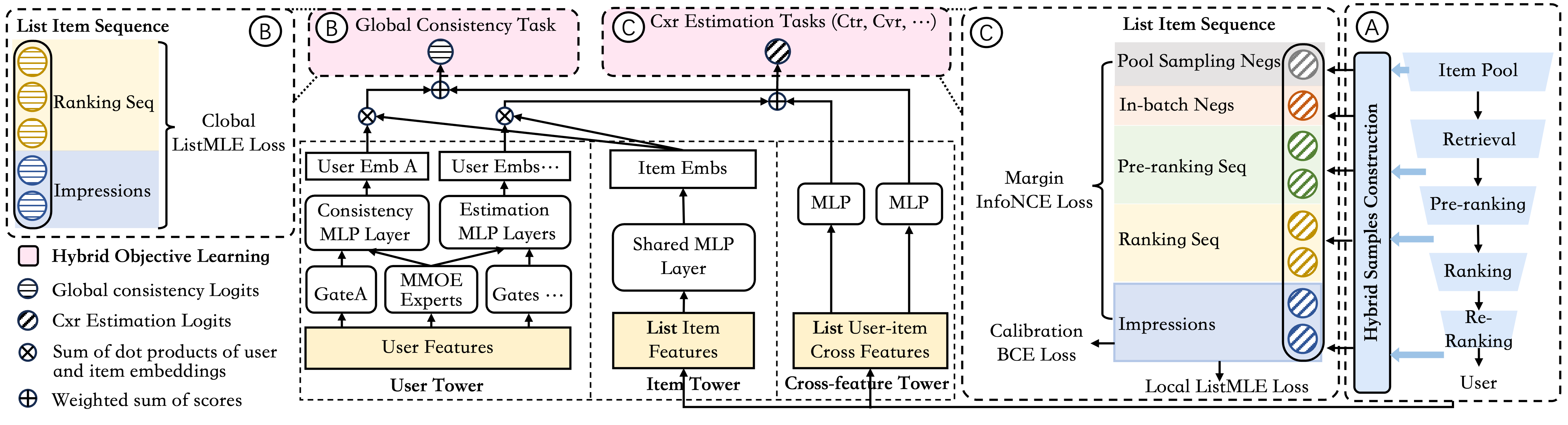}
\vspace{-8mm}
  \caption{Architecture of HCCP based on a three-tower model. (A) Hybrid sample construction module to obtain multi-level unexposed samples from upstream and downstream; (B) Consistency task to enhance the pre-ranking's rank ability on top items; (C) Hybrid optimization in cxr (ctr, cvr et.al) estimation tasks to improve the long-tail precision and mitigating SSB issue. (B) and (C) belong to hybrid objective learning, whose joint optimization achieves cross-stage coordination of pre-ranking.
}
\vspace{-3mm}
\label{fig:arch}
\end{figure*}

\vspace{-3mm}
\section{Related Work}
We briefly describe three aspects of research: pre-ranking, learning-to-rank algorithms, and sample selection bias issue.

\textbf{Pre-ranking}.
In industrial recommendation systems, pre-ranking plays a vital role in selecting high-quality items from large-scale retrieval items and relieving ranking pressure.
To balance effectiveness and efficiency, it widely adopts a vector-product-based deep neural network model\cite{2013dssm,2015lstmdssm}.
However, lightweight model architectures degrade the model expression ability. 
For critical calibration ability, recent works design model architectures to improve effectiveness under a computing power constraint, such as COLD\cite{2020cold}, FSCD\cite{2021fscd}, and IntTower\cite{2022inttower}.
Some works\cite{2020privileged,2020Meta-KD} focus on knowledge distillation. 
Despite some advancements, they optimize the pre-ranking model as an independent entity, neglecting the consistency with ranking, leading to sub-optimal outcomes.
For ranking ability, some  methods\cite{2022rankingConsistency,2023jdRethink} propose the importance of improving the consistency with ranking.
Some works propose strict score alignment via point-wise distillation loss\cite{2018_rank_distillation,2023rethinking}, MSE loss\cite{2022rankflow}, or ListNet distillation loss\cite{2023_list_distillation} to align with raw scores predicted by ranking. 
COPR\cite{2023_copr_ndcg} proposes a rank alignment module and a chunk-based sampling module to optimize the consistency with ranking. 
However, they treat all items equally, neglecting accuracy discrepancies between head and tail items predicted by ranking.
Moreover, pre-ranking has larger-scale candidates than ranking and the only utilization of ranking logs can not adapt to changes in retrieval distribution, leading to a sample selection bias problem.

\textbf{Learning-To-Rank (LTR) Algorithm}.
LTR algorithms are extensively used to enhance ranking abilities by approximating relevance degree, which can be categorized into point-wise, pairwise, and listwise.
Point-wise\cite{1994inferring} approaches usually treat ranking as regression or classification problems, which assign score for each item independently.
Considering mutual information, pairwise\cite{2005RankNet,2015RankingSVM} methods achieve local optimality by predicting relative orders of item pairs but ignoring the overall order.
Listwise methods\cite{2007onlylist,2008listmle1,2010lambdarank} consider the entire sequence quality from a global perspective, aligning with ranking objectives.
To preserve calibration and ranking quality, some studies combine the above losses\cite{2022point_mix2,2023joint_rank_cali}.

\textbf{Sample Selection Bias (SSB) Problem}.
As the magnitude of retrieval items increases, SSB issues\cite{2004ssb} in pre-ranking receive growing attention.
Some works sample negatives through a pre-defined distribution\cite{2013_global_neg} or in-batch sampling mechanisms\cite{2019LDRER,2020_negative_gold}.
Since the diversity of negatives sampled by in-batch methods is constrained by batch size, some works propose cross-batch sampling based on memory banks\cite{2020cross_cvpr,2021_cross_batch}.
In addition to negative sampling techniques, some approaches enhance the learning of long-tail items by cross-decoupling network\cite{2023ELIRCDN} or modified losses, such as the modified softmax loss integrated with multi-positive samples\cite{2023GCL_MOPPR}.
We propose an approach that leverages entire stream data to alleviate SSB issues and enhance adaptability when retrieval distribution shifts.

\vspace{-1mm}
\section{METHODOLOGY}
In this section, we present \textbf{H}ybrid \textbf{C}ross-Stage \textbf{C}oordination \textbf{P}re-ranking Model (\textbf{HCCP}).
As outlined in Section 3.2, we construct hybrid multi-level samples from the entire stream.
To leverage the strength of sampled unexposed information, we propose a hybrid object optimization approach in Section 3.3, designed for flexible adaptation across various pre-ranking models.

\vspace{-2mm}
\subsection{Overall Architecture}

\subsubsection{\textbf{Base Model}}
To balance efficiency and effectiveness, 
we design pre-ranking based on the Three-Tower\cite{2020cold} paradigm, which includes user, item, and cross-feature towers.
An MMOE\cite{2018mmoe} architecture is implemented on the user-side tower to facilitate multi-task learning.
Traditional multi-objective estimation tasks usually serve as calibration tasks, adopting binary cross-entropy (BCE) loss on impressions (exposed items).
The loss $ L_{cali}^{t}$ is defined as:
\vspace{-3mm}
\begin{equation}
    L_{cali}^{t} = - \frac{1}{|U|} \sum_{u \in U} \sum_{j=1}^{|R^{t}|} [ l^{t}_jlog(y_j^{t}) + (1-l^{t}_j) log( 1-y_j^{t}) ]
    \label{eq:calibration}
\vspace{-1mm}
\end{equation}
, where $y_j^{t}$ and $l_j^{t}$ are predicted score and corresponding label of item $j$ in task $t$. $R^t$ is the impression set and $U$ is the user request set.
Item embeddings are pre-computed and updated in near real-time.
Upon online request, user embedding and cross-feature scores are calculated in real-time. 
Scores are computed by dot-product across user and item embeddings, weighted with cross-feature scores.

\vspace{-2mm}
\subsubsection{\textbf{Innovation}}
To alleviate the SSB issue and improve precision, we employ a hybrid sampling technique that captures multi-level data from downstream and upstream stages of pre-ranking. 
Since unexposed samples are not associated with user feedback, we introduce an optimal hybrid objective optimization approach specifically tailored for such instances.
By explicitly integrating entire stream information, pre-ranking achieves cross-stage coordination and becomes a more suitable bridge between retrieval and ranking, instead of an isolated stage.
It can be seamlessly incorporated into any pre-ranking model with almost no latency increase.

\textbf{Hybrid Multi-level Sample Construction}.
To improve consistency, we sample \textbf{ranking sequence} ($S_2$ in Figure\ref{fig:cascade}) from downstream (ranking, re-ranking) stages in a non-uniform manner, where items ranked top have a higher sampling rate compared to lower.
Then the ranking sequence is rearranged based on user feedback of impressions (exposed items), which not only integrates ranking information but also the context and diversity information of re-ranking, and user interest.
From upstream, we collect three types of data: \textbf{pre-ranking sequence with latter order} in retrieved items ($S_1$ in Figure\ref{fig:cascade}), \textbf{in-batch} and \textbf{pool sampling negatives} from the item pool space ($S_0$ in Figure\ref{fig:cascade}).
Typically, since ranking and pre-ranking sequences are retrieved from upstream, they inherently exhibit a degree of alignment with user preferences.
Compared to in-batch and pool sampling negatives, they are usually considered hard negative samples.
The above samples improve the pre-ranking's discriminative capacity, especially when upstream provides a large number of long-tail items during online service.

\textbf{Hybrid Objective Optimization}.
Various multi-level samples serve different learning objectives. 
(1) \textbf{Learning consistency with downstream}: 
The ranking sequence offers unexposed information along with downstream feedback, which is crucial to improve the consistency with downstream.
We design list-wise global (on ranking candidate space) and local consistency (on exposure space) modules based on ranking sequence, as illustrated in Figure\ref{fig:arch}. 
Despite the constraints inherent in lightweight pre-ranking models, soft order alignment enables the pre-ranking to approximate the ranking ability of complex ranking models.
(2) \textbf{Optimizing long-tail precision throughout entire-stream data}:
Since hybrid unexposed data generally lack user feedback, we identify unexposed negatives and potential positives through contrastive learning. 
To incorporate prior knowledge of negative difficulties into learning, we introduce the Margin InfoNCE loss. 
Its efficacy in selecting potential positives is elaborated upon in the appendix.
While enhancing long-tail item precision, our method concurrently mitigates the SSB issue.

\vspace{-2mm}
\subsection{Hybrid Multi-level Sample Construction}
Except for clicked or purchased items as positives and un-clicked or un-purchased as negatives in impressions (N1), we collect various samples from downstream and upstream stages by online serving and offline sampling to enhance precision and alleviate SSB issues:
\begin{itemize}[leftmargin=*]
\vspace{-4mm}
\item N2, unexposed ranking sequence $B^{r}=\{j|j\in R^c, e_j=0\}$. $R^c$ is non-uniformly sampled from downstream and then arranged based on exposure information and user feedback. 
\item N3, pre-ranking sequence $B^{p}$ with latter orders, 
uniformly sampled from retrieved candidates($S_1$), but not entering downstream.
\item N4, in-batch negatives $B^{i}$, randomly sampled from impressions paired with other users in the same batch\cite{2020_negative_gold}.
\item N5, pool sampling negatives $B^{c}$, randomly sampled from the predefined item pool space ($S_0$) but not been retrieved.
\end{itemize}
We delineate the multi-level sample construction methods as follows.
The final dataset is organized in a \textbf{list-wise} format to ensure that samples in one request are learned in the same batch.
The list-wise manner can enhance model throughput and training efficiency (Table\ref{tab:complexity}) and improve the diversity of unexposed in-batch negatives.



\begin{figure}
  \centering
  \includegraphics[width=0.44\textwidth, height=3.0cm]{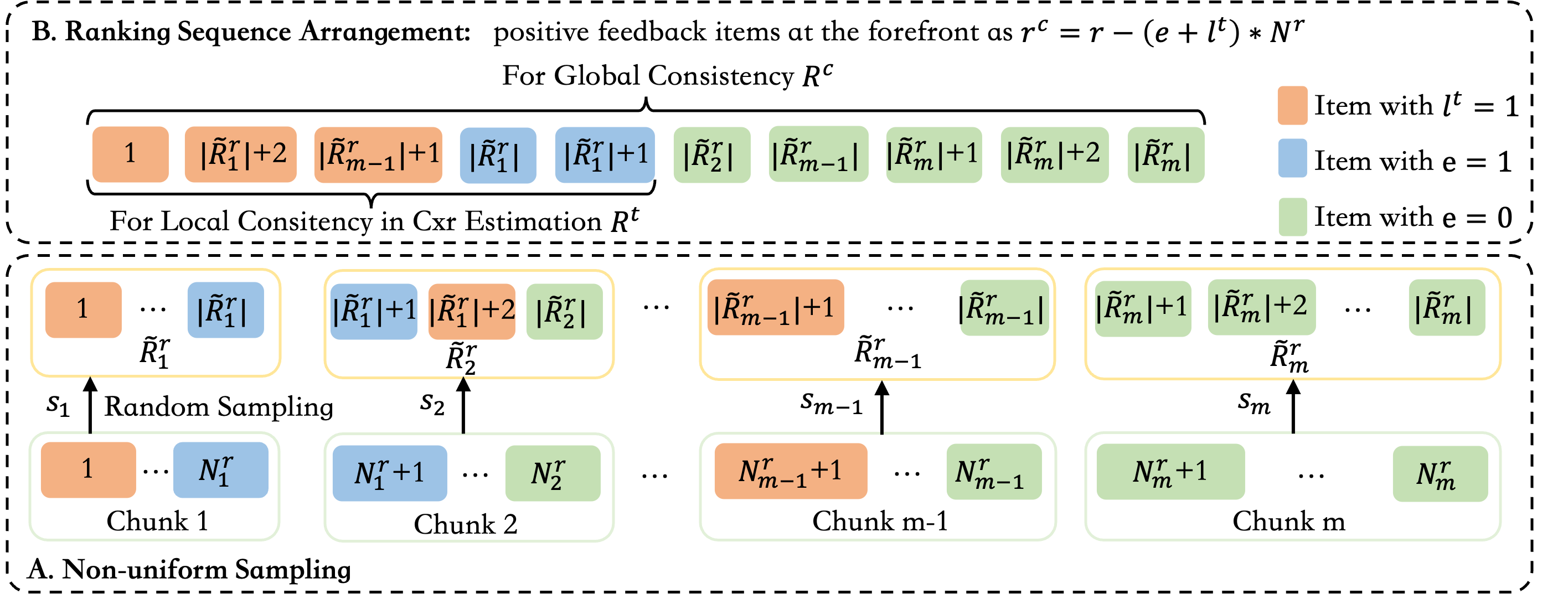}
  \vspace{-3mm}
  \caption{Ranking Sequence Construction. 
  (A) Non-uniform sampling: randomly sampling items within uneven chunks.
  (B) Sequence arrangement for global and local consistency.
  }
  \vspace{-5mm}
\label{fig:sample}
\end{figure}

\vspace{-1mm}
\subsubsection{\textbf{Ranking Sequence}}
The ranking sequence provides feedback information predicted by downstream stages, which can help to enhance consistency and precision.
To construct the optimal guiding "ground truth" ranking candidates conducive to pre-ranking training, we implement a non-uniform sampling paradigm and arrange them according to exposure information and user feedback.

\textbf{Non-uniform Sampling Strategy}. 
Ranking models typically optimize the precision of impressions and neglect tail items\cite{2017tail,2020tail}, leading to long-tail items being lower, yet they may align with user interests.
Directly learning scores or orders predicted by ranking cannot achieve optimum efficiency due to inaccurate predictions of tail items\cite{2023rethinking}.
Thus, it is necessary to prevent long-tail bias and ensure pre-ranking's accuracy for long-tail items is not affected by orders predicted by ranking.
Considering accuracy discrepancies between top and tail items of ranking, we design the non-uniform sampling strategy assigning higher sampling rates to high-ranked (top) items and lower rates to low-ranked (tail) items. 
In each request of user $u$, pre-ranking delivers top $N^r$ items to ranking.
We denote $R^r$ as sorted ranking sequence with ranking orders.
$$
    R^r = [(u, r_1, e_1, \textbf{l}_1), ... , (u, r_{N^r}, e_{N^r}, \textbf{l}_{N^r})], \ (r_1<...<r_{N^r})
$$
, where $r$ is the ranking order, $e$ and $\textbf{l}$ are one-hot exposure and a vector of task labels, such as click or purchase $l^{clk}, l^{ord}$.
$l^{clk} = 1$ or $e = 1$ mean item is clicked or exposed, while $l^{clk} = 0$ and $e = 0$ indicate the opposite. Items with smaller $r$ are sorted to the front.
\begin{figure}[h]
  \centering
  \includegraphics[width=0.45\textwidth, height=3.2cm]{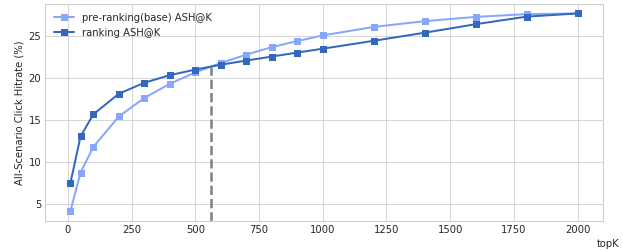}
  \vspace{-5mm}
  \caption{ The all-scenario click hitrate (ASH@K) of ranking and pre-ranking in the JD recommendation system.
}
  \vspace{-4mm}
\label{fig:fig_hit_rank}
\end{figure}
Then we map the original ranking sequence to $m$ chunks with different sample rates $s_j$, which is illustrated in Figure\ref{fig:sample} and Eq~\ref{eq:sampling_rate}.
\begin{equation}
s_j = \left\{ \begin{array}{ll}
                    s_1 & \text{if } 1 \leq r_j < N_1^r, \\
                    \vdots & \quad \vdots \\
                    s_m & \text{if } N_m^r \leq r_j \leq N^r
                    \end{array} \right.
    \label{eq:sampling_rate}
\end{equation}
, where $r_j$ is the predicted ranking order of item $j$. 
It is adaptable to various mapping functions.
The sampled ranking candidates are recorded as $\widetilde{R}^r$, which keep $M=\sum_i^m s_i\cdot N^r_i$ samples.
Noticeably, impressions are retained during sampling.
Sample rates control the consistency granularity and are flexibly adjusted according to online logging pressure.
As $s_j$ increases, consistency shifts from coarse-grained to fine-grained.
It ensures increased sampling for top items and reduced sampling for tails while preserving randomness.
For instance, in JD recommendation scenario, the pre-ranking's ASH@K (All-Scenario Click Hitrate@K)\cite{2023rethinking} exceeding ranking as $K$ exceeds 600, as shown in Figure \ref{fig:fig_hit_rank}.
Thus, a threshold of 600 can be established, whereby items ranked above it are assigned a higher sampling rate, while those below it receive a lower sampling rate.

\textbf{Sequence Arrangement}. 
To mitigate the bias of raw ranking sequences, we rearrange sampled sequence $\widetilde{R}^r$ to align pre-ranking learning requirements.
Exposure and click-task labels reflect the item selection tendency of downstream and user interests.
Impressions incorporate a broader range of diversity and contextual information.
As shown in Figure\ref{fig:sample}, we define the arranged ranking sequence $R^c$ for the global consistency task as follows.
$$
    R^c = [(u, r^c_1, e_1, \vec{l}_1), ..., (u, r^{c}_M, e_M, \vec{l}_M)] 
$$
, where $R^c$ is sorted in ascending order according to adjusted order $r^c = r-(e+l^{clk})*N^r$. 
It preserves global order prioritizing clicked items first, followed by unclicked impressions, and finally unexposed items, while also maintaining the local order within the three parts as per the original sampled sequence $\widetilde{R}^r$.
Additionally, to increase local ranking ability, we integrate ranking sequence orders with different task labels in exposure space.
The order of exposure sequence $R^t=[j|j \in R^c, e_j=1]$ in Cxr estimation tasks is defined as $r^t = r-(e+l^{t})*N^r, \ (t \in \{clk, ord, ...\})$.

\subsubsection{\textbf{Pre-ranking Sequence with Latter Order}}
Retrieval stages provide $N^p$ items to pre-ranking, in which top $N^r$ items are delivered downstream.
We denote $R^p$ as pre-ranking sequence with latter pre-ranking order.
$$
    R^p = [(u, \textbf{l}_1), ..., (u, \textbf{l}_j)... , (u, \textbf{l}_{N^p-N^r})], \ (1 \leq j \leq N^p-N^r)
$$
It captures items not entering the ranking and exposed.
Although they belong to hard negatives, they may still contain a wealth of information and match user interests, including potential positives to some extent.
The sampled pre-ranking sequence $B^p$ is randomly sampled from the original sequence $R^p$ as a fixed sample rate.
They are usually discarded in previous pre-ranking training.
However, we can incorporate them into training through contrastive learning.
It improves the ability to distinguish hard negatives especially when upstream stages provide long-tail items during online service.

\subsubsection{\textbf{In-batch Negatives}}
We sample unexposed negatives by employing a list-wise in-batch sampling method within impressions of other users.
They usually act as easy negatives to achieve popular item suppression, which is described as the "gold negative" in \cite{2020_negative_gold}.

\begin{figure}
  \centering
\vspace{-1mm}
  \includegraphics[width=0.45\textwidth, height=2cm]{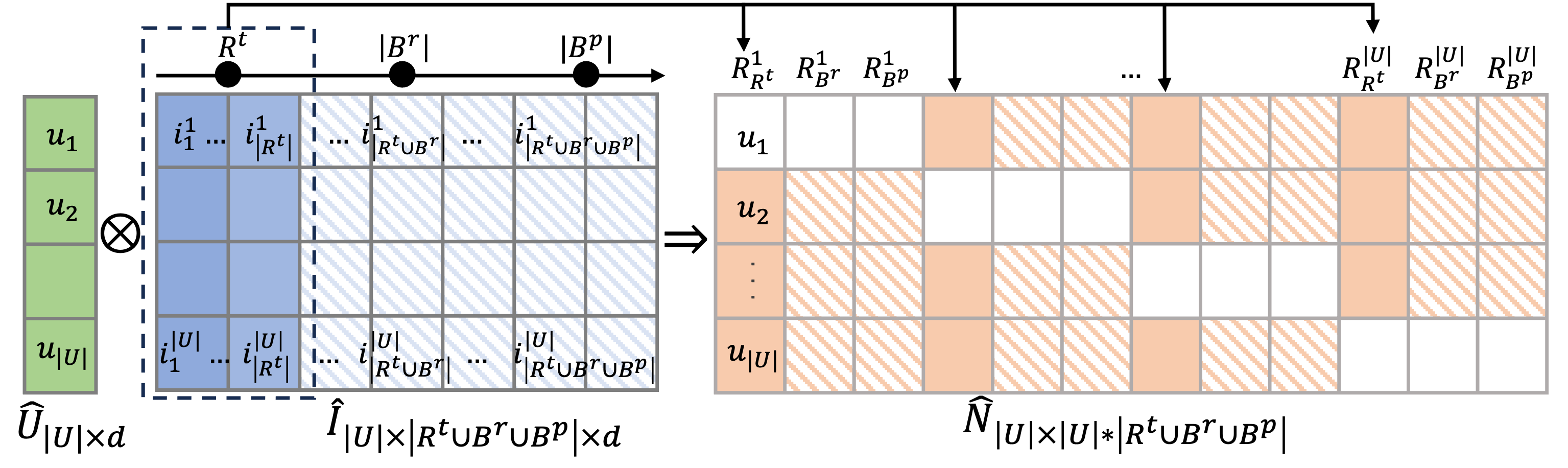}
  \vspace{-4mm}
  \caption{List-wise in-batch sampling. Top and tail items paired with other users can be sampled in different ratios.
  }
  \vspace{-5mm}
\label{fig:neg}
\end{figure}

The initial set to be scored is the intersection of impressions $R^t$, ranking $B^r$ and pre-ranking negatives $B^p$.
As illustrated in figure\ref{fig:neg}, for estimation task $t$, the list-wise in-batch negative sampling is achieved as following steps:
(1) Reshape predicted item embeddings $\hat{I}$ into a shape of $(|U|*|R^t \cup B^r \cup B^p|, d)$, where $|U|$ is request numbers in a batch and $d$ is the embedding dimension.
(2) Apply a transpose operation on $\hat{I}$ to the shape of $(d, |U|*|R^t \cup B^r \cup B^p|)$.
(3) Perform a cross product on user embedding $\hat{U}_{|U| \times d}$ and item embedding $\hat{I}_{d \times |U|*|R^t \cup B^r \cup B^p| }$ to obtain negative score tensor $\hat{N}_{|U| \times |U|*|R^t \cup B^r \cup B^p| }$.
(4) For user $u$, mask negative scores that belong to corresponding item collection $R^t, B^r, B^p$ and sample final in-batch negatives $B^{i}$ from $\hat{N}$.
To reduce the computational complexity, the cross product is calculated on item embedding $\hat{I}_{|U| \times |R^t| \times d}$ of exposure sequence $R^t$.
Compared to traditional in-batch sampling, whose diversity is determined by batch size $|U|$, while the magnitude of negatives in list-wise in-batch sampling is $|U| \times |R^t|$, which helps to improve the diversity of unexposed negatives.

\vspace{-1mm}
\subsubsection{\textbf{Pool Sampling Negatives}} 
Such negatives $B^c$ are randomly sampled from the item pool space, usually containing items that have not been retrieved, which can be regarded as easy negatives.
When upstream expands the retrieval magnitude and provides massive unseen items when online serving, introducing this kind of negatives can improve the pre-ranking's discriminative ability.

\vspace{-1mm}
\subsection{Hybrid Objective Optimization}
\subsubsection{\textbf{Consistency with Downstream Learning}}
To improve pre-ranking's consistency with downstream, we design a global consistency task on ranking sequence space and a local consistency module on exposure space. 
It achieves a soft alignment with downstream, which helps to obtain the ranking ability of complex ranking models, further improving the precision.

\textbf{(1) Global Consistency}:
It enables pre-ranking to optimize the ranking ability by approximating downstream from a \textbf{global perspective}.
In Figure\ref{fig:arch}B,
the global consistency loss is designed based on ListMLE loss\cite{2008listmle1} to explicitly learn ranking sequence order $R^c$, which provides consideration of global orders while pair-wise loss considers only relative order. 
The loss $L^c$ is defined as:
\vspace{-3mm}
\begin{equation}
    L^c = - \frac{1}{|U|} \sum_{u \in U} \sum_{j=1}^{|R^c|} log \frac{exp(y_j^{c})}{\sum_{k=j}^{|R_{c}|} exp(y_k^{c})} 
    \label{eq:listmle1}
\vspace{-1mm}
\end{equation}
, where $y^{c}_j$ is the predicted score of item $j$ in global consistency task. 
Minimizing $L^c$ makes each item in the sequence $R^c$ have the highest top-one probability in the subsequence with latter order.

\textbf{(2) Local Consistency in Cxr Estimation}:
For Cxr estimation tasks (Ctr, Cvr, ...), we integrate traditional calibration task with sub-task local consistency, as shown in Figure\ref{fig:arch}C.
Expect the calibration loss $L^t_{cali}$ in Eq~\ref{eq:calibration} for ensuring calibration capability, we design the local consistency loss to emphasize ranking ability from a \textbf{local perspective} in exposure space.
The corresponding loss $L^{t}_{cons}$ ensures the local order alignment on impressions to enhance the ranking ability based on the exposure sequence $R^t$.
\vspace{-3mm}
\begin{equation}
    L_{cons}^{t} = - \frac{1}{|U|} \sum_{u \in U} \sum_{j=1}^{|R^t|} log \frac{exp(y_j^{t})}{\sum_{k=j}^{|R^t|} exp(y_k^{t})}  
\vspace{-2mm}
\end{equation}
Similar to global consistency loss defined in Eq~\ref{eq:listmle1}, minimizing $L^{t}_{cons}$ makes each item in exposure sequence $R^t$ have the highest top-one probability in the subsequence with latter order.


\subsubsection{\textbf{Long-tail Precision Optimization}}
In cascade recommendation systems, given that downstream input is derived from the pre-ranking output,
relying only on the downstream will limit the pre-ranking model, which may further aggravate the Matthew effect. 
Since pre-ranking serves as a bridge between retrieval and ranking stages,
it is necessary to strengthen the ranking ability and precision of mid and long-tail items and provide high-quality items to downstream.
Expect N1, samples N2 \textasciitilde N5 extracted from the entire stream can alleviate the SSB issue but typically lack user feedback.
To leverage these unexposed samples, we introduce an effective approach for negative learning from the perspective of contrastive learning. 
We propose the \textbf{Margin InfoNCE loss} to incorporate priors about hard or easy negatives into learning, which helps identify unexposed potential positives and negatives.
As illustrated in Figure\ref{fig:arch}C, cxr estimation tasks integrate traditional calibration and local consistency with negative contrastive learning.
It enhances pre-ranking's adaptability to upstream and guarantees robust discriminative capacity when upstream provides massive long-tail items. 
In the Appendix, we provide an analysis of how the proposed loss distinguishes potential positives from hard negatives.

\textbf{Margin InfoNCE Loss}. 
Different categories of negatives above vary in their levels of difficulty. 
Group N1 is usually top-ranked downstream and has a definite selection tendency from users, we directly utilize the BCE loss on them, as described in Section 3.1.1.
Unexposed samples in N2 \textasciitilde N3 are retrieved due to their relevance with the user, but they are ranked lower by downstream or pre-ranking models.
Although they have not been exposed, they contain potential positive samples.
Introducing these \textbf{hard negatives} helps to improve the ability to distinguish difficult samples.
Negatives in N4 \textasciitilde N5 typically lack direct relevance or interest alignment with users, classified as \textbf{easy negatives}.
Although they are both easy negatives, there are subtle differences in effects: introducing N4 can suppress popular items and relieve the Matthew Effect, while N5 introduces massive unseen items, which helps improve discrimination ability, especially when retrieval magnitude increases.

Compared to hard negatives that may contain potential positives, easy negatives are less aligned with user preferences, hence, their distance from positives needs to be sufficiently large.
Some traditional losses do not explicitly differentiate between types of negatives. 
Inspired by InfoNCE\cite{2018InfoNCE} and Margin Softmax\cite{2016LSoftmax,2017ssoftmax,2018AMSoftmax} losses that explicitly encourage discriminative feature learning to reduce intra-class variation and increase inter-class difference, we propose the Margin InfoNCE loss to explicitly differentiate between positives, hard (N2 \textasciitilde N3), and easy negatives (N4 \textasciitilde N5), as defined:
\vspace{-1mm}
\begin{equation}
    L_{neg}^{t} =  -\frac{1}{|U|} \sum_{u \in U} \sum_{j=1}^{|B^{t}|} log \frac{e^{ \frac{v_j^{t} \phi(\theta_j)}{\tau}}}{ e^{ \frac{v_j^{t} \phi(\theta_j)}{\tau}} + \sum_{i=1}^{|B^h|} e^{\frac{v_i^{t} \phi(\theta_i)}{\tau}} + \sum_{i=k}^{|B^e|} e^{\frac{v_k^{t} cos\theta_k}{\tau}}  
    \label{eq:margin_infonce}
    }  
\end{equation}
, where $\tau$ is the temperature value.
$B^{t} = \{ j | j\in R^t, l^t_j =1 \}$ contain positive-label items in impressions of task $t$.
$B^h=B^{r} \cup B^p$ is the union of ranking and pre-ranking negatives.
$B^e=B^{i} \cup B^{c}$ is the union of in-batch and pool sampling negatives.
$v_j^{t} =\|\mathbf{u^t}\| \|\mathbf{i_j}\|$ is the product of user and item embeddings.
The angle between user and item embedding is defined as 
$ cos\theta = \frac{\mathbf{u} \cdot \mathbf{i} }{\|\mathbf{u^t}\| \|\mathbf{i_j}\|} $.
Referring to AM-Softmax\cite{2018AMSoftmax}, we introduce the additive margin $\phi(\theta) = cos\theta-m, (m > 0)$ based on InfoNCE loss. 
In addition, we analyze why the Margin InfoNCE loss can distinguish potential positive items from hard negatives in the Appendix.
The overall loss is defined as:
\vspace{-1mm}
\begin{equation}
    L = \lambda_c L^c + \sum_t^T \lambda_t ( L_{cali}^{t} + \alpha L_{cons}^{t} + \beta L_{neg}^{t} )
\end{equation}
, where $\lambda_c, \lambda_t, \alpha, \beta$ are loss weight of global consistency, multi-objective estimation, local consistency, and negative tasks.

\vspace{-1mm}
\section{EXPERIMENT}
We conduct a series of offline experiments on public Taobao\footnote{https://tianchi.aliyun.com/dataset/dataDetail?dataId=56} and JD production datasets and online A/B testing to evaluate the effectiveness of our method.
We aim to answer the following questions:

\begin{itemize}[leftmargin=*]
\item \textbf{Q1}: Does HCCP enhance the overall performance and serve as a suitable bridge between retrieval and ranking stages?
\item \textbf{Q2}: How does each learning objective perform in performance?
\item \textbf{Q3}: Does HCCP have an advantage over previous methods in mitigating the mid and long-tail bias issue?
\item \textbf{Q4}: Does the complexity of HCCP adapt to industrial systems?
\end{itemize}





\begin{table*}
  \caption{Comparison of Overall Performance of Various Methods on the JD Production Dataset}
  \vspace{-3mm}
  \label{tab:jd_result1}
  \begin{tabular}{p{0.1cm}p{1.8cm}|p{0.57cm}p{0.60cm}p{0.60cm}p{0.60cm}|p{0.57cm}p{0.60cm}p{0.60cm}p{0.60cm}|p{0.57cm}p{0.6cm}p{0.6cm}p{0.60cm}|p{0.60cm}p{0.57cm}p{0.60cm}p{0.60cm}}
    \toprule
    \multicolumn{2}{c|}{\multirow{2}{*}{Method}} & \multicolumn{4}{c|}{ISH@K (pp)}&\multicolumn{4}{c|}{ISPH@K(pp)} & \multicolumn{4}{c|}{ASH@K(pp)}&\multicolumn{4}{c}{ASPH@K(pp)}  \\
    \multicolumn{2}{c|}{} &10 & 100 & 1000 & 2000 &10 & 100 & 1000 &2000 &10 & 100 & 1000 & 2000 &10 & 100 & 1000 & 2000\\
    
    \midrule
    \multirow{2}{*}{G1} & \texttt{Base} & - & - & - & - & - & - & - & - & - & - & - & - & - & - & - & - \\
    \multirow{2}{*}{} &\texttt{Base+ListNet}   & 0.09 & -0.26 & -1.48 & -0.98   & -0.17 & -0.67 & 3.42 & 3.59     & 0.07 & 0.33 & 0.46 & 0.40       & -1.41 & 0.59 & 2.92 & 3.53   \\
    \midrule
    \multirow{1}{*}{G2} &\texttt{Base+ListMLE} & 0.20 & 1.17 & 0.19 & 0.19     & 0.50 & -0.22 & 3.84 & 4.51     & 0.03 & 0.35 & 1.17 & 1.07      & 0.41 & 2.11 & 3.98 & 5.56\\
    \midrule
    \multirow{7}{*}{G3} &\texttt{COPR}  & 1.56 & 8.19 & 14.10 & 9.69     & 1.67 & 8.79 & 27.31 & 27.69            & 1.60 & 7.21 & 16.18 & 14.03       & 0.70 & 8.34 & 23.42 & 24.69 \\
    \multirow{7}{*}{} & \texttt{Rethink}& 4.49 & 16.10 & 19.05 & 11.20   & 6.67 & 29.74 & 51.20 & 38.66    & 2.48 & 9.00 & 16.74 & 13.96        & 3.87 & 21.12 & 31.69 & 29.36 \\
    \multirow{7}{*}{} &\textbf{\texttt{HCCP(Ours)}}             & \textbf{9.09} & \textbf{22.21} & \textbf{21.31} & \textbf{12.56}    & \textbf{16.86} & \textbf{40.07} & \textbf{54.59} & \textbf{40.23}       & \textbf{14.32} & \textbf{26.43} & \textbf{25.96} & \textbf{18.85}      & \textbf{12.07} & \textbf{24.52} & \textbf{38.68} & \textbf{34.19} \\
    \hdashline
    \multirow{7}{*}{} &\texttt{w/o Up}     & 1.63 & 7.90 & 15.38 & 11.44     & 1.34 & 7.95 & 29.99 & 29.36       & 0.80 & 3.94 & 12.93 & 11.67       & 0.47 & 8.42 & 23.37 & 22.70    \\
    \multirow{7}{*}{} &\texttt{w/o PRC}    & 2.56 & 11.65 & 19.34 & 13.33    & 2.51 & 13.30 & 41.44 & 35.46      & 1.50 & 7.02 & 14.92 & 12.86     & 0.70 & 11.86 & 24.05 & 26.29\\
    \multirow{7}{*}{} &\texttt{w/o Neg}    & 6.46 & 20.27 & \textbf{22.94} & \textbf{14.01}    & 12.12 & 34.71 & 54.40 & 39.23     & 5.69 & 15.84 & 22.36 & 17.44      & 6.50 & 23.70  & 34.05 & 32.06\\
    \multirow{7}{*}{} &\texttt{w/o Margin} & 8.49 & 21.76 & 21.22 & 12.89    & 14.69 & 39.23 & 54.12 & 39.57     & 10.71 & 22.16 & 24.36 & 18.09      & 10.10 & 23.33  & 34.28 & 30.08\\
    \multirow{7}{*}{} &\textbf{\texttt{HCCP(Ours)}}             & \textbf{9.09} & \textbf{22.21} & 21.31 & 12.56    & \textbf{16.86} & \textbf{40.07} & \textbf{54.59} & \textbf{40.23}       & \textbf{14.32} & \textbf{26.43} & \textbf{25.96} & \textbf{18.85}      & \textbf{12.07} & \textbf{24.52} & \textbf{38.68} & \textbf{34.19} \\
    \bottomrule
  \end{tabular}
   \vspace{-1mm}
   \begin{tablenotes}
     \item[1] Due to the data sensitivity, the table only displays the absolute increase value.
   \end{tablenotes}
   \vspace{-3mm}
\end{table*}

\begin{table}
  \caption{Performance Comparison on Public Taobao Dataset}
  \vspace{-3mm}
  \label{tab:res_taobao}
  \begin{tabular} {p{1.75cm}|p{0.72cm}|p{0.75cm}|p{0.77cm}|p{0.75cm}|p{0.77cm}|p{0.6cm}}
    \toprule
    Method & AUC & Impr & ISH@2 & Impr & ISH@5 & Impr  \\
    \midrule
    \texttt{Base}         &0.7515  &-  &0.4947 &- & 0.7531 &-   \\
    \texttt{Base+ListNet}    &0.7684  &1.70pp  &0.4991 & 0.45pp & 0.7580 & 0.50pp \\
    \texttt{Base+ListMLE}    &0.7669  &1.54pp  &0.4973 &0.27pp &0.7568 &0.37pp   \\
    \texttt{COPR}            &0.7713  &1.98pp  &0.4986 &0.39pp &0.7567 &0.36pp    \\
    \texttt{Rethink}         &0.7791  &2.77pp  &0.5102 &1.55pp &0.7621 &0.90pp    \\
    \textbf{\texttt{HCCP(Ours)}}  &\textbf{0.7807} &\textbf{2.92pp} &\textbf{0.5151} &\textbf{2.04pp} &\textbf{0.7675}&\textbf{1.44pp}   \\
    \bottomrule
  \end{tabular}
  \vspace{-4mm}
\end{table}

\vspace{-2mm}
\subsection{Experiment Setup}
\subsubsection{Dataset} 
\textbf{Public Taobao Dataset} contains 26 million expression logs in 8 days.
We utilize DIN\cite{2018DIN} as the ranking model to guide the learning of pre-ranking methods.
Note that the public dataset lacks multi-level unexposed data, causing a significant gap between simulation and industrial systems for methods Rethink\cite{2023rethinking} and ours.
\textbf{JD Production Dataset} contains sampled data from the entire stream of clicked requests in one week on the JD homepage. 
The request amount for training is the magnitude of billions.
The validation is conducted on pre-ranking candidates from one day that occurred one week later.

\vspace{-2mm}
\subsubsection{Metrics}
The model performance is evaluated on the metrics.

\textbf{Offline metrics}: 
1) Traditional AUC metrics reflect ranking abilities on impressions (top items). Considering that the public dataset contains only impressions, we use AUC to evaluate the performance.
Due to inherent bias in historical impressions, AUC is difficult to align with online performance.

2) Since pre-ranking aims at distilling an unordered optimal subset, 
we follow in-scenario click/purchase hitrate (\textit{ISH}, \textit{ISPH}) and all-scenario click/purchase (\textit{ASH}, \textit{ASPH}) described in \cite{2023rethinking} to measure the quality of topK items selected by pre-ranking.
\vspace{-1mm}
$$
hit@K = \frac{1}{|U|} \sum_{u \in U} \frac{\sum_j^k \mathbf{1}(j \in I_u)}{|I_u|}
$$
\vspace{-1mm}
, where $\mathbf{1}$ is the indicator function to determine whether item $j$ in topK set is clicked and $I_u$ denotes the click or purchase sets of the current scenario and all scenarios when calculating ISH, ASH, ISPH and ASPH, respectively.
When calculating ASH/ASPH in recommendation systems, we directly use clicked and purchased items in all scenarios without query relevance filtering, which is different from search systems\cite{2023rethinking}.
ISH/ISPH better reflects users' interest in the current scenario, while ASH/ASPH reflects users' long-term and non-instant gratification interests.
Without focusing solely on top items, we set $K=10,100,1000,2000$ to evaluate the pre-ranking performance on top, middle, and tail items.

3) Mean average precision ($MAP@K$) and normalized discounted cumulative gain ($NDCG@K$)\cite{2009_ltr1} metrics can measure the consistency with ranking. 
However, consistency with ranking does not fully measure the performance of pre-ranking, which is analyzed in the experimental section.
In the JD production dataset, we use topK (K=10,100,1000) candidates selected by ranking as relative ground-truth values.
Due to the insufficient accuracy of ranking for tail items, we do not measure the tail set consistency (K=2000).


\textbf{Online metrics}:
In the JD E-commerce recommendation system, \textit{UCTR} (click PV/exposure UV) and \textit{UCVR} (purchase PV/exposure UV) are used for online performance evaluation.


\subsubsection{Baselines} Baseline methods are divided into 3 groups: \textbf{G1}, using only impressions; \textbf{G2}, using impressions with ranking orders; \textbf{G3}: using impressions and other unexposed samples.
And we design multi-versions of HCCP for ablation studies.
(1) \textbf{Base} (G1) adopts an MMoE architecture \cite{2018mmoe} and is trained on exposures by point-wise binary cross-entropy (BCE) loss.

(2) \textbf{Base+ListNet} (G1) implements a ranking and calibration joint optimization method through BCE classification loss and ListNet losses\cite{2007onlylist} based on the base MMoE model.

(3) \textbf{Base+ListMLE} (G2) employs listMLE \cite{2008listmle1} and BCE losses to improve ranking ability by training on impressions with ranking orders. 
It can be seen as HCCP with only local consistency tasks.

(4) \textbf{COPR}\cite{2023_copr_ndcg} (G3) implements a relaxation net, and CTR, pair-wise consistency, regularization losses using ranking candidates. 
We modify the underlying model to MMoE for a fair comparison.

(5) \textbf{Rethink}\cite{2023rethinking} (G3) achieves a list-wise loss on impressions, ranking, and pre-ranking candidates and a distillation loss on impressions with predicted CTR, which provides a hard consistency alignment.
This paper describes ASPH and online GMV drop when distillation extends to all ranking candidates.

(6) \textbf{HCCP} (G3) implements some different versions of HCCP. 
\begin{itemize}[leftmargin=*]
\item \textbf{HCCP(w/o Up)} improves consistency through a global consistency task, \textit{without N2\textasciitilde{}N5 samples} in estimation tasks.
\item \textbf{HCCP(w/o PRC)} achieves an InfoNCE loss only on in-batch negatives and pool sampling negatives \textit{without ranking and pre-ranking candidates N2\textasciitilde{}N3} based on HCCP(w/o Up).
\item \textbf{HCCP(w/o Neg)} uses the InfoNCE loss on candidates N2\textasciitilde{}N3 without \textit{sampled negatives N4\textasciitilde{}N5} based on HCCP(w/o Up).
\item \textbf{HCCP(w/o Margin)} uses InfoNCE loss \textit{instead of our margin InfoNCE loss} on N2\textasciitilde{}N5 samples. There is no explicit discrimination between N2\textasciitilde{}N3 (hard negatives) and N4\textasciitilde{}N5 (easy negatives).
\item \textbf{HCCP(Ours)} is the final version described in this paper.
\end{itemize}

\vspace{-2mm}
\subsubsection{Implementation Detail}
When constructing samples, we split ranking candidates into $7$ non-uniform chunks and set sample rates of different chunks as $0.5$, $0.125$, $0.1$, $0.05$, $0.025$, $0.01$, $0.005$ from top to tail.
In each request, the sampled ranking sequence $R^c$ and pre-ranking sequence $B^p$ contain $35$ and $15$ items with a $0.15\%$ sample rate.
In-batch sampling is implemented during model training, and its sampling rate is set to be $0.05$.
For pool sampling negatives, we sample 10 items as easy negatives in each request.

The number of experts in base MMoE model is set as 8.
Since it is difficult to obtain cross-features of N4\textasciitilde N5 negatives, such as the proportion of the item brand in the user click sequence, we directly use the InfoNCE loss to optimize the final weighted score of the Three Tower on N2\textasciitilde N3 samples.
User and item embedding dot-products are optimized on N2\textasciitilde N5 samples by the Margin InfoNCE loss.
The final loss is the average of the two losses.
Cxr estimation tasks contain click, purchase, and diversity tasks.
The diversity task regulates the item category diversity delivered to downstream stages.
The loss weight for overall global consistency, click, purchase, and diversity tasks are $0.05, 0.98, 0.2, 0.1$, respectively.
Within cvr estimation tasks, loss weights of local consistency $\alpha$ and margin InfoNCE losses $\beta$ are $0.05$ and $0.5$.
Margin and temperature values are $m=0.9$ and $\tau = 0.1$ in Margin InfoNCE loss.
Performance is somewhat sensitive to hyper-parameters, especially the global consistency's loss weight and the margin value in Margin InfoNCE loss, which are determined through ablation experiments.

\begin{table}
  \caption{Comparison of Consistency on JD Production Dataset}
  \vspace{-3mm}
  \label{tab:jd_result2}
  \begin{tabular}{p{2.5cm}|p{0.57cm}p{0.57cm}p{0.57cm}|p{0.57cm}p{0.57cm}p{0.57cm}}
    \toprule
    {Method} & \multicolumn{3}{c|}{MAP@K (pp)}&\multicolumn{3}{c}{NDCG@K (pp)}   \\
     &10 &100 & 1000 &10 &100 & 1000 \\
    \midrule
    \texttt{Base+ListNet}& 0.13 & 0.03 & -1.27 & 0.20 & -0.02 & -0.49 \\
    \midrule
    \texttt{Base+ListMLE}& 0.18 & 0.30 & -0.68 & 0.46 & 0.45 & -0.17 \\
    \midrule
    \texttt{COPR}        & 1.11 & 2.52 & \textbf{2.85} & 2.95 & 3.47 & \textbf{1.11} \\
    \texttt{Rethink}     & 1.77 & 2.17 & -2.44 & 3.96 & 3.16 & -0.83 \\
    \textbf{\texttt{HCCP(Ours)}}    & \textbf{4.06} & \textbf{4.29} & -0.11 & \textbf{7.60} & \textbf{5.63} & 0.01 \\
    \hdashline
    \texttt{HCCP(w/o Up)}   & 1.40 & 4.06 & 5.61 & 3.63 & 4.90 & 1.77 \\
    \texttt{HCCP(w/o PRC)}  & 1.61 & 3.93 & 5.54 & 3.98 & 4.94 & 1.87   \\
    \texttt{HCCP(w/o Neg)}  & 1.77 & \textbf{4.65} & \textbf{6.07} & 4.29 & 5.49 & \textbf{1.97}  \\
    \texttt{HCCP(w/o Margin)}& 4.00 & 4.11 & -0.48 & 7.46 & 5.46 & -0.15  \\
    \textbf{\texttt{HCCP(Ours)}}           & \textbf{4.06} & 4.29 & -0.11 & \textbf{7.60} & \textbf{5.63} & 0.01  \\
    \bottomrule
  \end{tabular}
  \vspace{-5mm}
\end{table}

\vspace{-2mm}
\subsection{Offline Performance Comparison (Q1)}
\subsubsection{Performance on Taobao Dataset}
Although introducing only impressions with ranking orders, HCCP achieves a slight improvement, as illustrated in Table\ref{tab:res_taobao}.
Since the public dataset lacks unexposed data, Table\ref{tab:res_taobao} can not fully reflect differences between Rethink and HCCP, we conduct experiments on JD production dataset.

\vspace{-6mm}
\subsubsection{Performance on JD Production Dataset}
As demonstrated in Table\ref{tab:jd_result1}, HCCP markedly surpasses other baseline methods on JD dataset.
The increase of the hitrate metrics when K=10 and 100 demonstrates HCCP's superior capability in identifying top items that closely align with user preferences.
When K=1000 and 2000, the highest hitrate metric of HCCP shows its higher prediction accuracy on mid and tail items.
The performance difference between Base and Base+ListNet is minimal, which suggests that directly introducing Listnet loss of learning-to-rank methods on original impressions cannot lead to significant improvements.
Additionally, we observe that the consistency of HCCP is highest on top items.
As shown in Table\ref{tab:jd_result2}, as K increases to 1000, both MAP and NDCG exhibit signs of decline, which is attributed to the fact that the optimization objective of HCCP is not consistency.
On the one hand, since ranking models predict top items more accurately, blindly improving the consistency with ranking on tail items may affect the pre-ranking accuracy on tail items.
On the other hand, since the role of pre-ranking is performing a preliminary selection from large-scale items, and distilling an optimal unordered subset in the cascading system, 
consistency affects pre-ranking performance to a certain extent, but it does not fully reflect the changes in pre-ranking performance.
Above analyses indicate that the optimization objective for pre-ranking should be to find a \textbf{balance} between \textbf{consistency with ranking} and \textbf{precision on mid and tail items}.

\begin{figure}[h]
  \centering
    \includegraphics[width=0.44\textwidth, height=2.5cm]{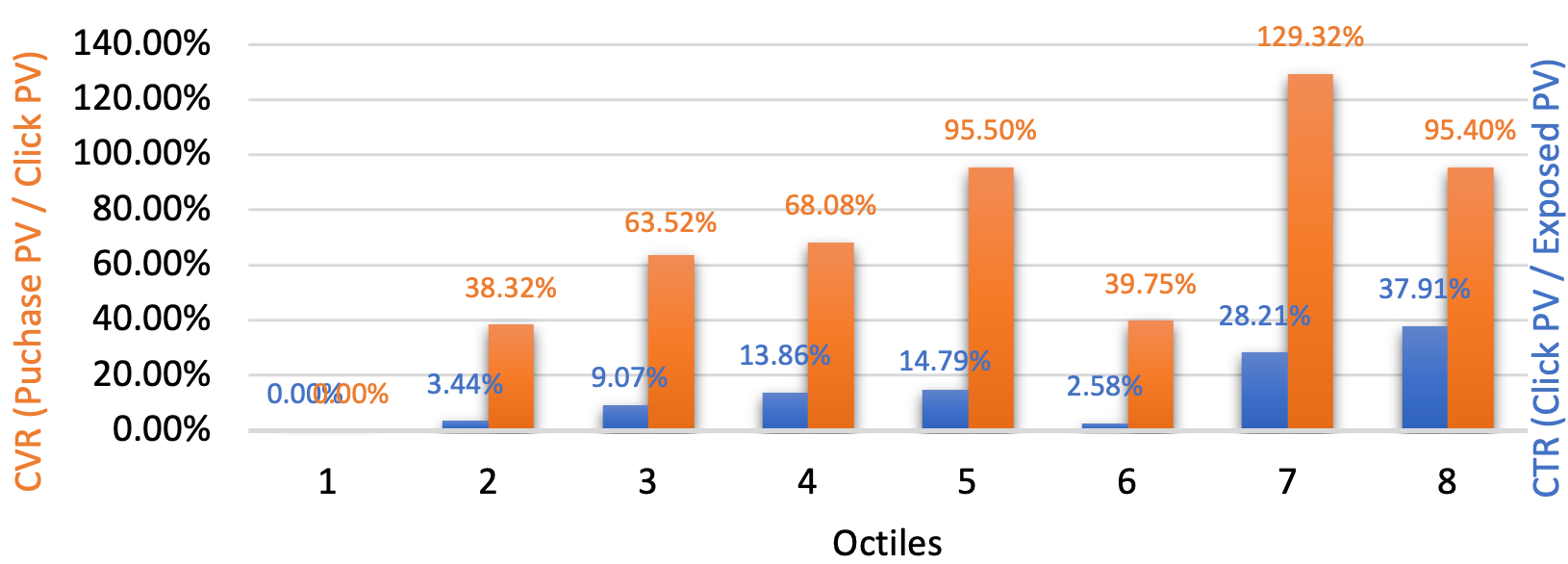}
  \vspace{-5mm}
  \caption{ CTR and CVR improvements of items with different frequencies compared to top items in HCCP. A larger quantile represents more long-tail items.
  }
  \vspace{-3mm}
\label{fig:cvr_ctr}
\end{figure}

\begin{figure}[h]
  \centering
    \includegraphics[width=0.46\textwidth, height=2.5cm]{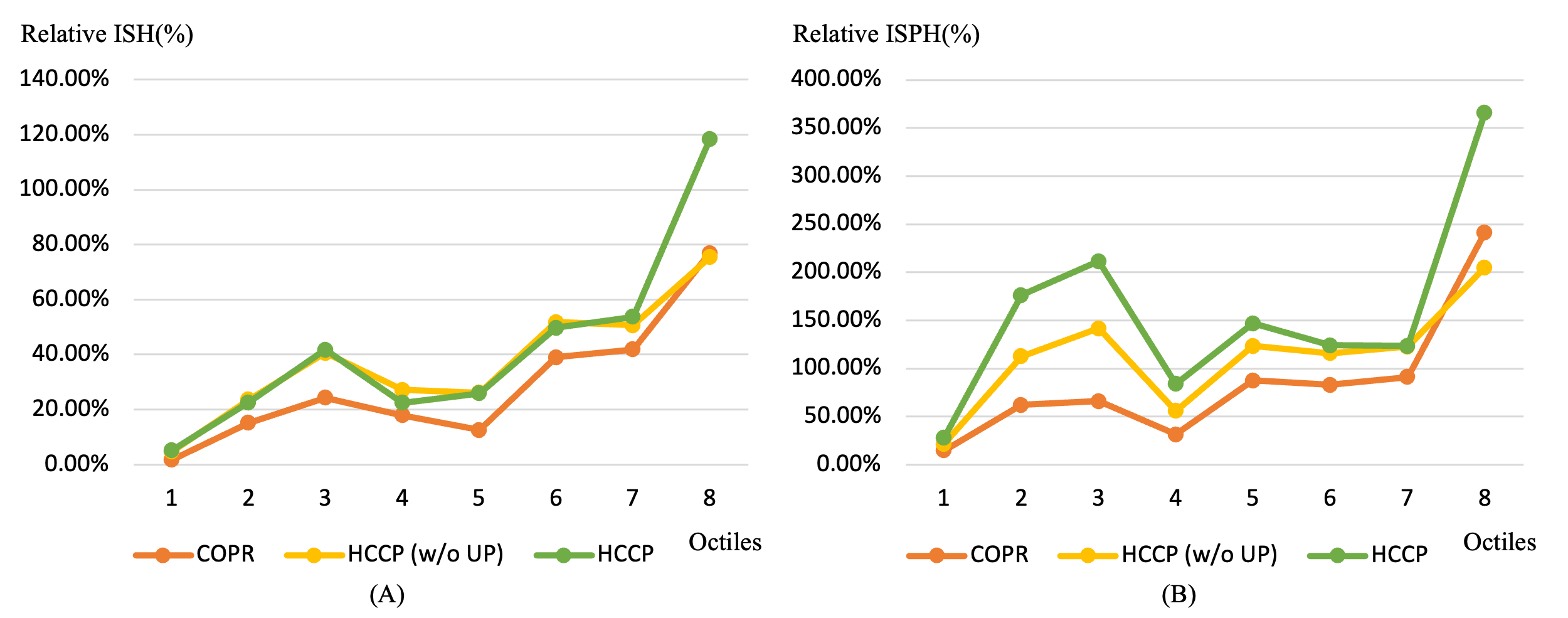}
  \vspace{-4mm}
  \caption{Relative ISH/ISPH@100 of HCCP, HCCP(w/o Up), and COPR on different frequency items. Quantiles from small to large represent transition from popular to long-tail.
  }
  \vspace{-4mm}
\label{fig:auc_tail_head}
\end{figure}

\vspace{-4mm}
\subsection{Online A/B Test}
We conduct the A/B test on JD homepage recommendation system for 7 days.
HCCP achieves a \textbf{14.9\%} increase in UCVR and \textbf{1.3\%} in UCTR, which is deployed online to serve main user traffic.
Among them, HCCP(w/o Up) with consistency modeling contributes up to 7.1\% UCVR and 0.6\% UCTR, and the inclusion of long-tail precision optimizing on multi-level samples provides an additional 7.8\% to UCVR and 0.7\% UCTR.
Moreover, we analyze the CTR and CVR improvement of items with different frequencies in Figure\ref{fig:cvr_ctr}.
When HCCP is deployed in JD system, there is a notable trend that mid and tail items have a significant improvement in CVR.
The phenomenon elucidates the more pronounced enhancement observed in UCVR, underscoring the efficacy of optimization long-tail precision.

\vspace{-2mm}
\subsection{Futher Analysis}
\subsubsection{Ablation studies (Q2)}
Ablation studies conduct to evaluate the effectiveness of each objective in joint optimization.
Results are displayed beneath the dashed lines in Table\ref{tab:jd_result1} and Table\ref{tab:jd_result2}.

(1) \textbf{The Importance of Local Consistency in Cxr Estimation Tasks}. 
Slight improvement on Base+ListNet compared with Base indicates that incorporating ranking orders of impressions (local consistency) can enhance pre-ranking accuracy to some extent.

(2) \textbf{The Importance of Global Consistency}. 
HCCP(w/o UP) trains global consistency on the ranking sequence of both exposed and unexposed data based on Base+ListNet, which achieves an enhancement across all metrics. 
It significantly outperforms COPR on consistency metrics, demonstrating the efficiency of ListMLE loss for order learning from a global perspective in learning-to-rank.

(3) \textbf{The Importance of Hybrid Samples in Long-tail Precision Optimization}. 
We validate the impact of easy (N4\textasciitilde N5) and hard (N2\textasciitilde N3) negatives through two experiments, designated as HCCP(w/o PRC) and HCCP(w/o Neg).
N2\textasciitilde N5 samples are important, however, directly incorporating them cannot maximize utility; for instance, HCCP(w/o Margin) leads to a performance or consistency decline on ISP@1000, ISPH@1000, and ASPH@2000, which is due to the introduction of hard negatives.
Thus we propose the Margin InfoNCE loss to distinguish these negatives explicitly.
However, more hard negatives increase the difficulty of learning long-tail items, which provides guidance for subsequent optimization.

(4) \textbf{The Importance of the Margin InfoNCE Loss}. 
We propose the Margin InfoNCE Loss to enhance the discriminative ability of negatives, where hard negatives in N2\textasciitilde N3 may contain potential positives and easy negatives in N4\textasciitilde N5 are less aligned with user interest.
Compared to HCCP(w/o Margin), HCCP better leverages unexposed data and outperforms on top, middle, and tail items.

\begin{table}
  \caption{Summary of training complexity and serving latency}
  \vspace{-3mm}
  \label{tab:complexity}
  \begin{tabular}{p{1.65cm}|p{0.88cm}|p{0.8cm}|p{2.6cm}|p{0.4cm}}
    \toprule
    Method & \#Param  & \#Data  & Offline Training & TP99 \\
    \midrule
    Base       & 6618w & 0.395B & $O(F_u(d)+F_i(d))$ & -  \\
    COPR     & 6622w & 2.757B & $O(F_u(d_1/r)+F_i(d_1))$ & - \\
    Rethink     & 6640w & 3.938B & $O(F_u(d_2/r)+F_i(d_2))$ & -  \\
    HCCP(Point) & 6641w & 4.747B & $O(F_u(d_3)+F_i(d_3))$ & - \\
    \textbf{HCCP(Ours)}  & 6641w & 4.747B & $O(F_u(d_3/r)+F_i(d_3))$ & +0.73\%\\
    \bottomrule
  \end{tabular}\vspace{-1mm}
  \begin{tablenotes}
     \item[1] B means "billion" in \#Data. TP99 is the 99th percentile of latency.
   \end{tablenotes}
  \vspace{-3mm}
\end{table}

\subsubsection{Performance on long-tail items (Q3)}
HCCP attains superior outcomes in hitrate metrics when K=2000 in Table\ref{tab:jd_result1}, indicating that it outperforms other methods in performance on tail items.
Moreover, we divide our validation dataset into octiles based on the frequency of item occurrences, with the first being popular items.
Figure\ref{fig:auc_tail_head} demonstrates that HCCP with multi-level samples significantly improves the mid and long-tail item precision while enhancing top item precision, effectively alleviating the seesaw phenomenon compared to HCCP(w/o Up) and COPR.
HCCP(w/o Up) and COPR exclusively emphasize consistency, which is demonstrably less effective in mitigating long-tail bias compared to HCCP.

\subsubsection{Complexity and Time Analysis(Q4)}
Base method utilizes the JD production dataset with only impression data, which contains 39.5 million instances.
After adding multi-level unexposed data without user feedback, the data volume reachs 4.75 billion.
Table\ref{tab:complexity} shows that the parameter count increase of HCCP compared to Base is mainly due to adding a global consistency task based on MMOE. 
Similarly, the increased parameter of Rethink\cite{2023rethinking} is also attributed to introducing an exposure task.
$F_u$ and $F_i$ represent the model complexity of the user tower and the item/cross-feature tower, respectively.
Offline training complexity is affected by several common factors, i.e., $r$ (\#request of training data), $d$ (\#base training data), $d_1$ (\#augmented data in COPR), $d_2$ (\#augmented data in Rethink), $d_3$ (\#augmented data in our method).
Compared HCCP(Ours) with HCCP(Point), although the training data is augmented, the list-wise organization manner reduces the offline training complexity.
During online serving, the latency (TP99) cost increase of 0.73\% caused by a slight parameter increase is acceptable in our system.

\vspace{-1mm}
\section{CONCLUSION}
We propose HCCP, a hybrid pre-ranking model to improve cross-stage coordination within the entire stream, jointly optimizing consistency and long-tail precision through multi-level sample construction and hybrid objective optimization modules.
It outperforms SOTA methods, contributing up to 14.9\% UCVR and 1.3\% UCTR within 7 days of deployment in the JD E-commerce recommendation system.
In future work, we will extend the training data to all scenarios with all-scenario positive feedback information.

\appendix
\section{Appendix}
\definecolor{commentcolor}{RGB}{0,128,0} 
\renewcommand{\algorithmiccomment}[1]{\hfill\textcolor{commentcolor}{#1}}
\algrenewcommand\alglinenumber[1]{\footnotesize\ttfamily#1:}
\begin{algorithm*}
\caption{A Tensorflow-style Pseudocode of our proposed Margin InfoNCE Loss.}
\vspace{-0.7mm}
\label{alg:example}
\begin{algorithmic}[1]
    \Function{Cal\_Margin\_InfoNCE\_Loss}{ }
    \State cos\_theta = logits $/$ (u\_norm $*$ i\_norm + $1e^{-10}$) \Comment{\# logits, u\_norm, i\_norm: [B*K], to calculate $cos_{\theta}$ in Eq\ref{eq:margin_infonce}}
    \State logits\_m = (cos\_theta - margin) $*$ (u\_norm $\cdot$ i\_norm + $1e^{-10}$) \Comment{\# margin: float, to get the logits with the margin $v \cdot \phi(\theta_k)$ in Eq\ref{eq:margin_infonce}}
    \State l\_update = (logits\_m + beta $*$ logits) / (1 + beta) \Comment{\# beta: float, using a weighted combination to maintain stability}
    \State beta $*=$ scale \Comment{\# scale: float, beta=9999 at the beginning of training. As iteration steps increase, beta becomes smaller.}

    \State l\_p = tf.reshape(tf.where(p\_mask), l\_update, tf.fill(tf.shape(p\_mask), neg\_min) ), [-1,1]) \Comment{\# p\_mask: [B*K] positive mask, neg\_min: float}
    \State l\_n\_hard = tf.where(n\_hard\_mask, l\_update, tf.fill(tf.shape(n\_hard\_mask), neg\_min) ) \Comment{\# n\_hard\_mask: [B*K] hard negative mask}
    \State l\_n\_easy = tf.where(n\_easy\_mask, logits, tf.fill(tf.shape(n\_easy\_mask), float(neg\_min)) ) \Comment{\# n\_easy\_mask: [B*K] easy negative mask}
        
    \State l\_n = tf.concat([tf.reshape(l\_n\_hard, [B,K]), tf.reshape(l\_n\_easy], [B,K]), axis=1) \Comment{\# B: batch\_size, K: sequence length. Get l\_n: [B,2*K]}
    \State l\_n = tf.reshape(tf.repeat(tf.reshape(l\_n, [B,1,K]), repeats=K, axis=1), [-1,2*K]) \Comment{\# repeat l\_n ([B*K,2*K]) to match l\_p([B*K,1]) }
    \State l\_all = tf.gather\_nd(tf.concat([l\_p, l\_n], axis=1), tf.where(p\_mask))  \Comment{\# get positive and corresponding negatives. l\_all: [len(pos), 2*K+1]}
            
    \State loss = -tf.math.log\_softmax( l\_all/t )[:,0] \Comment{\# t: float, temperature. The first column of l\_all is positive logit, while remaining is negative}
    \State \textbf{return} tf.reduce\_sum(loss)
    \EndFunction
\end{algorithmic}
\end{algorithm*}
We analyze the ability of the proposed Margin InfoNCE loss to discriminate potential positives from hard negatives.
In formal binary cross-entropy (BCE) calibration tasks, unclicked impressions that reflect a clear user's tendency not to choose are seen as negatives.
We calculate the derivative of BCE loss to the predicted score $y$.
\vspace{-1mm}
\begin{equation}
    L_{BCE} = - l_jlog(s_j) - (1-l_j) log( 1-s_j) 
\end{equation}
\vspace{-3mm}
$$
    s_j = sigmoid(y_j) = \frac{1}{1+e^{-y_j}}
$$
\vspace{-3mm}
\begin{align}
\frac{\partial L_{BCE}}{\partial y_j} = \frac{1}{1+e^{-y_j}} - l_j 
\end{align}
When $j$ is a negative sample, $l_n=0, \frac{\partial L_{BCE}}{\partial y_n} = \frac{1}{1+e^{-y_n}}$.
We calculate the derivative of the Margin InfoNCE loss to the predicted score $y$.
\vspace{-1mm}
\begin{align}
    L_{M} &= - log \frac{e^{ v_j \phi(\theta_{j})/\tau}}{ e^{ v_j \phi(\theta_{j})/\tau} + \sum_{i=1}^{|B^h|} e^{v_{i} \phi(\theta_{i})/\tau} + \sum_{k=1}^{|B^e|} e^{v_{k} cos{\theta_{k}}/\tau}   }  \\
    &= - log \frac{e^{ y_j /\tau}}{ e^{ \frac{y_j}{\tau} } + e^{ \frac{y_n-m( v_j-v_n)}{\tau}} + \sum_{i=1}^{|B^h|-1} e^{ \frac{y_i-m( v_j-v_i)}{\tau}} + \sum_{k=1}^{|B^e|} e^{\frac{y_{k}}{\tau}}  } \nonumber 
\end{align}
\vspace{-1.5mm}
\begin{figure}[h]
  \centering
  \vspace{-3mm}
\includegraphics[width=0.45\textwidth, height=4.4cm]{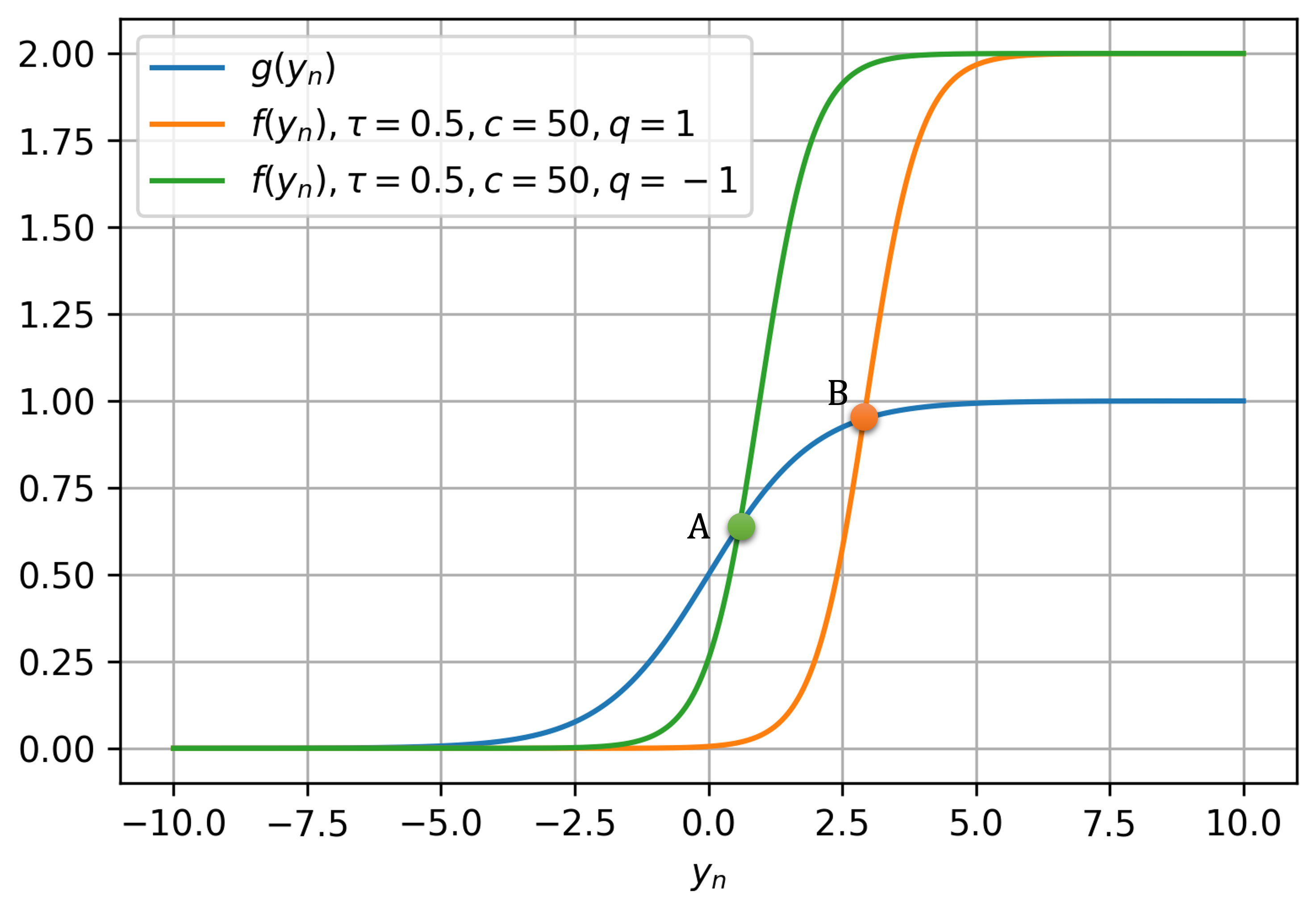}
  \vspace{-4.5mm}
   \caption{$f(y_n)$ and $g(y_n)$ functions.}
  \vspace{-5mm}
\label{fig:proof}
\end{figure}
$$
    v_j = \|\mathbf{u_j}\| \|\mathbf{i_j}\| \ , \ \ cos\theta_j = \frac{\mathbf{u_j} \cdot \mathbf{i_j}}{\|\mathbf{u_j}\| \|\mathbf{i_j}\|} \ , \ \  \phi_j = cos\theta_j - m
$$
, where $j$ and $n (n \in B^h)$ are the positive and negative items to be calculated the partial derivative. $B^h=B^r \cup B^e$ is the hard negative set and $B^e=B^i \cup B^c$ is the easy negative set. Since potential positives usually exist in hard negatives $B^h$, we only consider the gradient of positives $y_j$ and hard negatives $y_n (n \in B^h)$ here.
\begin{align}
\frac{\partial L_{M}}{\partial y_{n}} &= \frac{1}{\tau}  
 \frac{e^{ (y_{n} - m(v_j - v_n)) /\tau}}{ e^{ \frac{y_j}{\tau} } + e^{ \frac{y_n - m(v_j-v_n)}{\tau} } + \sum_{{i}=1}^{|B^h|-1} e^{\frac{y_i - m(v_j-v_i)}{\tau} } + \sum_{{k}=1}^{|B^e|} e^{\frac{y_k}{\tau}}  }  \\ \nonumber
&= \frac{1}{\tau} \frac{1}{ 1 + \sum_{i=1}^{|B^h \cup \{j\} |-1} e^{\frac{y_{i}-y_n+m(v_i-v_n)}{\tau}} + \sum_{k=1}^{|B^e|} e^{\frac{y_{k}-y_n+m(v_j-v_n)}{\tau}}  }  
\end{align}
When the same negative sample $y_n$ is optimized using different loss functions $L_{BCE}$ and $L_M$, the corresponding gradients can be approximated as $g(y_n)$ and $f(y_n)$ as Eq\ref{eq:grad}, respectively. 
We show them with different parameters in Figure\ref{fig:proof}, where $q$ and $c$ control the left and right movement of the entire $f(y_n)$ function along the x-axis, and $\tau$ and $c$ control the function's slope.
\vspace{-2mm}
\begin{align}
g(y_n) =  \frac{1}{ 1 + e^{(-y_n)}  } \ , \ \ f(y_n) = \frac{1}{\tau} \frac{1}{ 1 + c \cdot e^{(q-y_n)/\tau }  } 
\label{eq:grad}
\end{align}
As the value of the negative sample $y_n$ approaches $0$ (and includes subsequent steps), gradient $f(y_n)$ will be smaller than $g(y_n)$, meaning that the rate of descent will slow down. 
Different $q$, $c$, and $\tau$ will only affect the position of this dividing point (like A and B in Figure\ref{fig:proof}), but will not affect overall trend.
When optimizing, using our proposed Margin InfoNCE loss $L_M$ instead of the BCE loss $L_{BCE}$ will result in higher scores for potential positives in hard negative samples, thereby effectively selecting the potential positives.

\bibliographystyle{ACM-Reference-Format}
\bibliography{sample-base}

\end{document}